\documentclass[useAMS, usenatbib, usegraphicx]{mn2e}
\usepackage{times}

\title[The outbursts of RS~Ophiuchi]{A pre-outburst signal in the long term optical lightcurve of the recurrent nova  RS~Ophiuchi}

\author[S.~Adamakis et al.]{S.~Adamakis$^{1,2}$, S.~P.~S.~Eyres$^1$, A.~Sarkar$^{1,3}$, R.~W.~Walsh$^1$\\
$^1$Jeremiah Horrocks Institute, University of Central Lancashire,
Preston, PR1 2HE, UK\\
$^2$Decision Science, Lloyds Banking Group, 155 Bishopsgate, London,
EC2M 3TQ, UK\\
$^3$Space Science Center, University of New Hampshire, Durham, NH 03824, US\\}

\date{Accepted .  Received ; in original form}

\begin{document}
\label{firstpage}
\maketitle

\begin{abstract}
  Recurrent novae are binary stars in which a white dwarf accretes
  matter from a less evolved companion, either a red giant or a
  main-sequence star. They have dramatic optical brightenings of
  around 5--6~mag in V in less than a day, several times a
  century. These occur at variable and unpredictable intervals, and
  are followed by an optical decline over several weeks, and activity
  from the X-ray to the radio.  The unpredictability of recurrent
  novae and related stellar types can hamper systematic study of their
  outbursts. Here we analyse the long-term lightcurve of
  RS~Ophiuchi, a recurrent nova with six confirmed outbursts, most
  recently in 2006~February. We confirm the previously suspected 1945
  outburst, largely obscured in a seasonal gap. We also find a signal
  via wavelet analysis that can be used to predict an incipient
  outburst up to a few hundred days before hand. This has never before
  been possible. In addition this may suggest that the preferred
  thermonuclear runaway mechanism for the outbursts will have to be
  modified, as no pre-outburst signal is anticipated in that case.  If
  our result indeed points to gaps in our understanding of how
  outbursts are driven, we will need to study such objects carefully
  to determine if the white dwarf is growing in mass, an essential
  factor if these systems are to become Type Ia
  Supernovae. Determining the likelihood of recurrent novae being an
  important source population will have implications for stellar and
  galaxy evolution.
\end{abstract}


\section{Introduction}
\label{sec-intro}

The binary star RS~Ophiuchi (RS~Oph) consists of a white dwarf (WD)
orbiting within the dense wind of a red giant (RG). As the WD travels
in its orbit it accretes matter from the RG wind, growing slowly in
mass. The orbital period is around 460~days (Dobrzycka and Kenyon,
1994), but the system does not eclipse as the inclination angle is
approximately 50$\circ$ to the line-of-sight, assuming a massive WD
around 1.2 - 1.4 solar masses (Brandi, 2009). It has experienced at
least six dramatic optical brightening events or outbursts (in 1898,
1933, 1958, 1967, 1985 and 2006) generally attributed thermonuclear
runaway (TNR) events within the accreted matter on the WD surface. Two
other events are suggested in the literature \citep[1907; 1945
  from][respectively]{Schaefer2004,Oppenheimer1993}. The optical
development during an outburst is very similar in each case
\citep{Rosino1986}. Following the 2006 outburst, a WD mass of around
1.35 solar masses has been suggested \citep{Sokoloski2006} implying
that it may be a Supernova~Ia progenitor.

Since the WD is gaining mass through accretion, but losing mass due to
the outbursts, an interesting question is: what happens to the overall
mass of the WD after an outburst? Will it increase, stay unchanged or
decrease over many outbursts? If it actually increases in mass then
this means that at some point it will exceed the Chandrasekhar limit,
which will lead to a Type~Ia Supernova
\citep{Sokoloski2006}. Supernovae outbursts play an important role in
the history of the universe and as standard candles. Being able to
observe a complete recurrent nova cycle from pre-outburst accretion to
post-outburst quiescence in sufficient detail will tell us a lot about
the physical processes involved. If we were able to predict an
outburst, we would be able to take spectroscopic and multi-band
observations of the star before and at the beginning of the outburst,
which has never been done before.

Here we examine the optical lightcurve between 1933 and 2008, which
includes five confirmed and one suspected outburst. The data have been
taken from the American Association of Variable Star Observers
\citep[][AAVSO hereafter]{Henden2009} and are mainly visual
estimates. In Section~\ref{sec-lightcurve} we present the lightcurve
over 75~years and discuss previous work to analyse the variations. In
Section~\ref{sec-decline} we look at a Bayesian approach to
parameterising the form of the lightcurve during the outbursts,
describing the stages of post-outburst development and providing a
model that supports the subsequent analysis of the whole
lightcurve. Section~\ref{sec-wavelet} presents a wavelet analysis of
the lightcurve, which leads to a pre-outburst signal that is present
in the data even if we remove the outbursts themselves.  In
Section~\ref{sec-discussion} we discuss this feature and its potential
utility in predicting outbursts sufficiently early to allow detailed
observations of the pre-outburst and peak periods of the next
outburst. 

\section{Optical lightcurve}
\label{sec-lightcurve}

Fig.~\ref{fig-lc} shows the AAVSO visual lightcurve from 1933
August 16th up until the 2008 January 31st (44,655 points over
27196~days). These are mainly naked eye visual estimates from amateur
astronomers, while more recently these are supplemented by V band
telescope measurements, also from amateur observers. There is
considerable scatter, but the data is of sufficient quality to
determine some basic features:
\begin{figure*}
\includegraphics[angle=0, width=18.5cm]{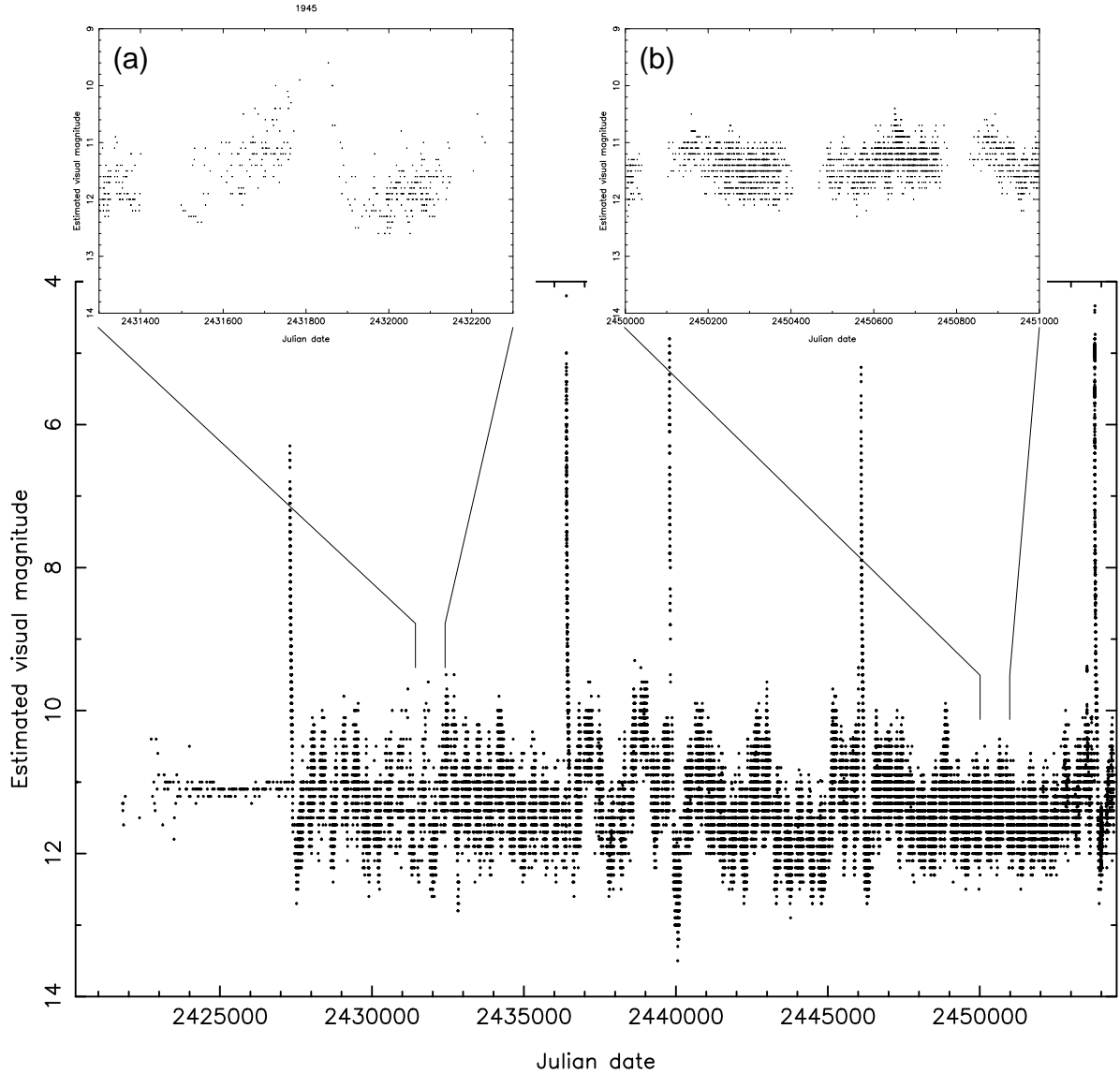}
\caption{Optical lightcurves from 1933 to 2008, mainly visual
  estimates from the AAVSO. Panel (a) shows the 1000 days centred on
  1945 (see Fig.~\ref{fig-outbursts}) while (b) shows the 1000 days
  centred on 1997, showing a typical quiescent period including two
  seasonal gaps. The horizontal ``stripes'' are due to the data being
  recorded to only the nearest 0.1~mag on the AAVSO database.}
\label{fig-lc}
\end{figure*}

\begin{itemize}
\item
There are clear peaks in 1933, 1958, 1967, 1985 and 2006 due to the
optical brightening -- this is the fundamental activity that identifies
the object as a recurrent novae.
\item
The quiescent light curve fluctuates between 9.6 and 12.8
magnitudes, but with no periodicity or other pattern.
\item
The time from outburst to decline to quiescent mean magnitude is about
100 days, but the light consistently declines below this point and
only recovers to approximately the mean after 400 to 500~days.
\item
The seasonal gap from mid-November until late-January, is long enough
to contain most of an outburst, so that we may have missed one or
more. However each decline-and-recovery phase is sufficiently similar
and lasts long enough that the tail of such a hidden outburst might be
identifiable.
\end{itemize}

\begin{figure*}
\includegraphics[angle=0, width=18.5cm]{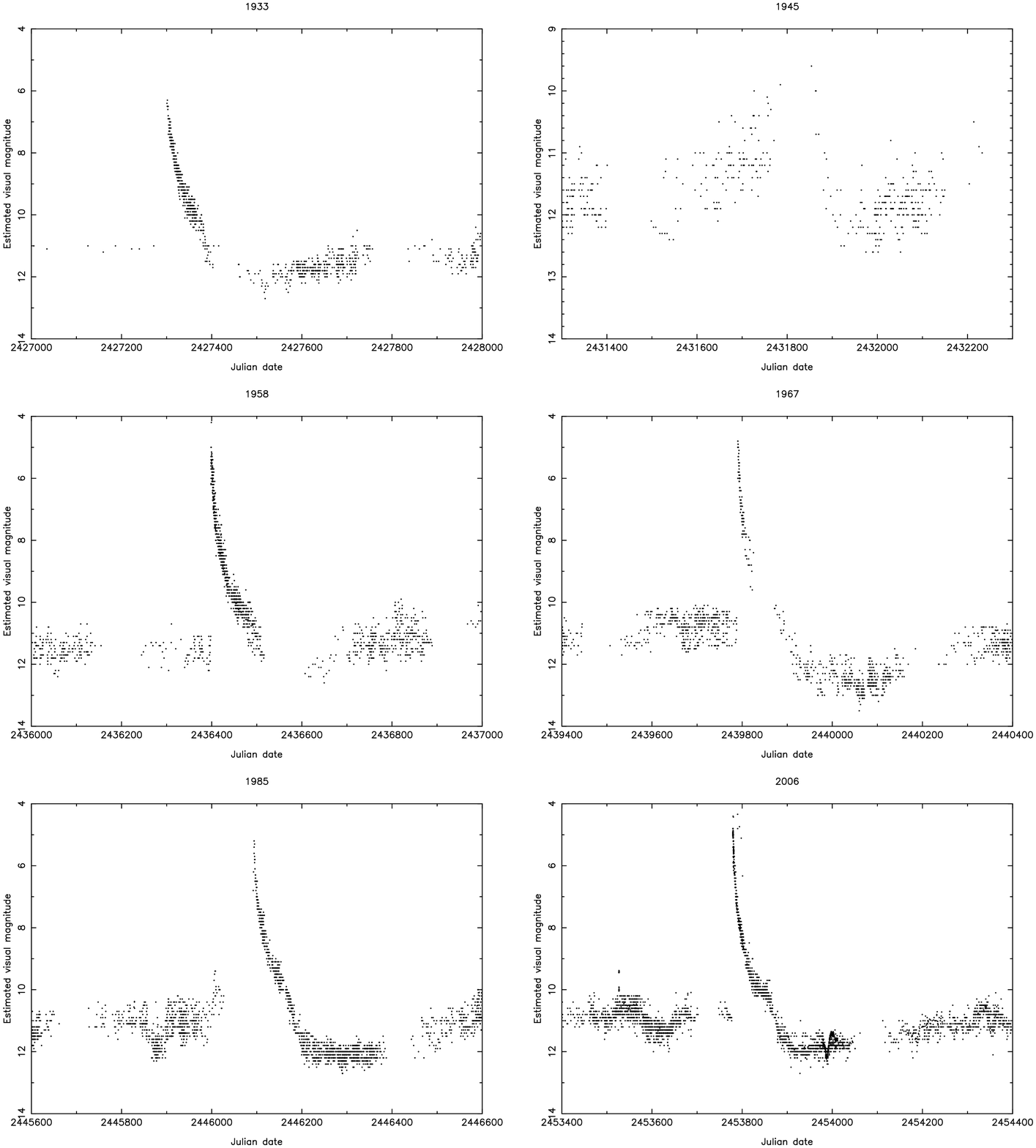}
\caption{Five confirmed outbursts present in Fig.~\ref{fig-lc} in the
  years as marked, plus the period of time in 1945 that
  \citet{Oppenheimer1993} propose shows the final decline and
  recovery phases of a sixth outburst. This panel is on a different
  vertical scale, and is also presented in Fig.~\ref{fig-lc} panel (a)
  to enable comparison with a typical quiescent period in panel
  (b). Also apparent in these panels is the increase density of data
  with time over the 75 years presented.}
\label{fig-outbursts}
\end{figure*}

As we examine the characteristics of the outbursts, we also present
the 1000~days around each outburst in Fig.~\ref{fig-outbursts}. We
also include the data around 1945 that led \citet{Oppenheimer1993} to
suggest an outburst that was missed in the seasonal gap.

A number of authors \citep{Dobrzycka1996,Sokoloski2001,Gromadzki2006,
  Worters2007,Zamanov2010} have found flickering--like fluctuations on
timescales of minutes, and \citet{Worters2007} were the first to
detect the resumption of flickering post--outburst. As flickering of
this sort is generally attributed to accretion in binary star systems,
this enabled the first post--outburst accretion rate estimate to be
made. The AAVSO lightcurve presented here has insufficient time
resolution to allow flickering analysis. However it should be noted
that night-to-night variations of 0.5~mag are seen in high time
resolution data, so that the AAVSO data have some sensitivity to the
underlying amplitude of flickering.

We extract the outburst events from Fig.~\ref{fig-lc} and conduct a
Bayesian analysis to characterise the decline, described in
Section~\ref{sec-decline}. We also use the data in Fig.~\ref{fig-lc}
as the basis of a wavelet analysis with a Morlet mother function
\citep{Torrence1998}, as described in Section~\ref{sec-wavelet}.

\section{Decline of the light curve following an outburst}
\label{sec-decline}

\citet{Oppenheimer1993} discussed the decline rate of the
outbursts. They divided each outburst into different phases according
to the decline rate and concluded that each outburst consists of three
breakpoints before the star begins to brighten again. In this section,
we model the outburst data with simple curves. The aim is simply to
parameterise the form of the decline, rather than determine any
underlying physical model.

In order to apply the statistical analysis, we take the first 500~days
of each outburst (including the proposed 1945 one). The total number
of data values is n = 8505. The most well observed outburst is the one
that happened in 2006 with 4109 observed points. The more observed
points there are in an outburst, the more information will be added
from this outburst to the statistical analysis we
undertake. \citet{Oppenheimer1993} commented that the decline rate and
then brightening rate of the proposed 1945 (obscured) outburst is akin
to all the other outbursts, and that this combination has never been
observed outside of an outburst. From this resemblance of the light
curves, we have a reason to believe that the same physical mechanisms
cause all the outbursts. Thence, we can assume that all the outbursts
can be described with the same parametrised model. 

\subsection{Models for Bayesian analysis}
\label{ssec-models}

We apply Bayesian analysis to the declines from maximum for each
outburst in the data. We consider four models. In each case the time
$t_4 = 500$~days defines the end of the outburst, and the parameters
$\gamma_1, \gamma_2, \gamma_3, \gamma_4$ are calculated to ensure each
model is continuous. We take time zero as a parameter to determine the
start of each outburst, instead of assuming it is defined by the first
observed point in outburst; this allow us to estimate the start date of
the proposed 1945 outburst and hence check that it would not have been
visible before the seasonal gap. The break points in the decline are
calculated as times $t_1, t_2$ and $t_3$ from this zero point.

{\bf Model 1.} This is the most complicated model of the four with 11 free
parameters. It consists of four stages. During the first three there
is a decline in magnitude, whereas in the last one it increases.
\begin{equation}
M_1: \mu = \left\{
\begin{array}{llll}
\gamma_1 + \alpha_1\log(t+\beta_1), & \quad 0 \le t < t_1 \\
\gamma_2 + \alpha_2\exp(\beta_2 t), & \quad t_1 \le t < t_2 \\
\gamma_3 + \alpha_3\exp(\beta_3 t), & \quad t_2 \le t < t_3 \\
\gamma_4 + \alpha_4 t, & \quad t_3 \le t < t_4 \\
\end{array}
\right.
\end{equation}
The parameter space for this model will be $\mathbf{P_1}=(\gamma_1,
\alpha_1, \beta_1, t_1, \alpha_2, \beta_2, t_2, \alpha_3, \beta_3,
t_3, \alpha_4)$.

{\bf Model 2.} This is the second most complex model with 10 free
parameters. It consists of four stages. The only difference between
this and the $M_1$ model is the third stage, where instead of an
exponential function we assume a straight line.
\begin{equation}
M_2: \mu = \left\{
\begin{array}{llll}
\gamma_1 + \alpha_1\log(t+\beta_1), & \quad 0 \le t < t_1 \\
\gamma_2 + \alpha_2\exp(\beta_2 t), & \quad t_1 \le t < t_2 \\
\gamma_3 + \alpha_3 t, & \quad t_2 \le t < t_3 \\
\gamma_4 + \alpha_4 t, & \quad t_3 \le t < t_4 \\
\end{array}
\right.
\end{equation}
The parameter space for this model will be $\mathbf{P_1}=(\gamma_1,
\alpha_1, \beta_1, t_1, \alpha_2, \beta_2, t_2, \alpha_3, t_3,
\alpha_4)$.

{\bf Model 3.} 
This model consists of nine free parameters and four stages. The only
change from the $M_2$ model is the second stage, where instead of an
exponential function we assume a straight line.
\begin{equation}
M_3: \mu = \left\{
\begin{array}{llll}
\gamma_1 + \alpha_1\log(t+\beta_1), & \quad 0 \le t < t_1 \\
\gamma_2 + \alpha_2 t, & \quad t_1 \le t < t_2 \\
\gamma_3 + \alpha_3 t, & \quad t_2 \le t < t_3 \\
\gamma_4 + \alpha_4 t, & \quad t_3 \le t < t_4 \\
\end{array}
\right.
\end{equation}
The parameter space for this model will be $\mathbf{P_1}=(\gamma_1,
\alpha_1, \beta_1, t_1, \alpha_2, t_2, \alpha_3, t_3, \alpha_4)$.

{\bf Model 4.} 
This is the simplest model with seven free parameters. The basic difference
between this and the previous three models is that it consists of
three stages. In the first two the magnitude is declining and in the
third it is increasing.
\begin{equation}
M_4: \mu = \left\{
\begin{array}{llll}
\gamma_1 + \alpha_1\log(t+\beta_1), & \quad 0 \le t < t_1 \\
\gamma_2 + \alpha_2 t, & \quad t_1 \le t < t_2 \\
\gamma_3 + \alpha_3 t, & \quad t_2 \le t < t_3 \\
\end{array}
\right.
\end{equation}
The parameter space for this model will be $\mathbf{P_1}=(\gamma_1,
\alpha_1, \beta_1, t_1, \alpha_2, t_2, \alpha_3)$.

In the following analysis, we assume that the first observed brightest
point of each outburst is at time $\tau_j, \quad j=1,\ldots,6$ from
the true beginning of the outburst (there are 6 outbursts including
the 1945 one). We incorporate $\mathbf{P_2}=(\tau_1,\ldots,\tau_6)$ as
parameters so that the data will decide their values. This will also
give us the information about how many days after the 1945 peak the
first observation occurred. Therefore, the parameter space will be
$\mathbf{P} = (\mathbf{P_1},\mathbf{P_2})$.

Uninformative prior distributions were assumed for the parameters,
where possible. Also normal and uncorrelated errors were implemented
for the observed values. In this case nuisance parameters like the
standard deviation of the errors can be integrated out.

\subsection{Preferred model}
\label{ssec-preferred}

First, we will try to choose which of the four models fits best to the
data we have and after that we present the parameter estimations for
that particular model. The codes used for this analysis were written
and implemented by \citet{Adamakis2009}.

Table \ref{M_j:BF} depicts the logarithmic marginal densities
estimation with three different approximation methods. All of them
choose model $M_2$ as the best, although $M_2$ is ``not more than a
bare mention'' better than $M_3$, according to the criteria of
\citet{Kass1995}. The information criteria (Table \ref{M_j:IC}) choose
$M_2$ and $M_3$ as the best two models. However, they disagree as AIC
suggests $M_2$, whereas BIC proposes $M_3$. The maximum likelihood
functions under these two models are $-26235.67$ for $M_2$ and
$-26237.19$ for $M_3$. Although $M_2$ maximises the likelihood
compared to $M_3$, the fact that it contains one more parameter than
$M_3$ does not allow Bayes factor estimations to clearly favour
$M_2$. Therefore, $M_3$ cannot be excluded. Nevertheless, we select
model $M_2$ for further analysis.

\begin{table}
\caption[Logarithmic marginal densities estimation for modelling
  RS~Oph's outbursts.]{Logarithmic marginal densities estimation for
  modelling RS~Oph's outbursts. 1: Laplace method with posterior
  covariance matrix, 2: Laplace method with robust posterior
  covariance matrix, 3: Importance sampling estimation with the
  probability density from stage $1$ as the additional probability
  density.}
\label{M_j:BF}
\begin{center}
\begin{tabular}{ccccc}     
  \hline                   
  & $M_1$ & $M_2$ & $M_3$ & $M_4$ \\
  \hline
1 & -26276.07 & -26271.55 & -26272.58 & -26301.42 \\
2 & -26277.63 & -26272.27 & -26273.36 & -26302.03 \\
3 & -26279.96 & -26274.97 & -26275.29 & -26301.33 \\
  \hline
\end{tabular}
\end{center}
\end{table}

\begin{table}
\caption{Information criteria and maximised log-likelihood functions
  for the $M_j,$ $j=1, \ldots, 4$ models.}
\label{M_j:IC}
\begin{center}
\begin{tabular}{ccccc}     
  \hline                   
  & $M_1$ & $M_2$ & $M_3$ & $M_4$ \\
  \hline
AIC & 52505.30 & 52503.35 & 52504.38 & 52568.91 \\
BIC & 52625.12 & 52616.12 & 52610.11 & 52660.54 \\
maximum & -26235.65 & -26235.67 & -26237.19 & -26271.46\\
log-likelihood & & & & \\
  \hline
\end{tabular}
\end{center}
\end{table}

Table \ref{M_1:estimation} provides all the information we need for
the parameter estimation. We will not comment separately on each
parameter because there are too many. Instead we will try to focus on
the most important ones. Starting with the break point parameters, we
can estimate $t_1$ to be $\sim 53$ days from the actual beginning of
the outburst (a $95\%$ probability credible interval will give it to
be between $\sim 49$ and $\sim 59$ days). Between $\sim 53$ days and
$\sim 107$ days ($95\%$ credible interval between 102 and 109) from
the beginning of the outburst, the decline rate changes from
logarithmic to exponential. The decline is completed $\sim 143$ days
after the beginning of the actual outburst ($95\%$ credible interval
between $\sim 139$ and $\sim 146$ days). The decline rates can be
taken from parameters $\alpha_1$, $\beta_1$, $\alpha_2$, $\beta_2$,
$\alpha_3$ and $\alpha_4$, with the $alpha$ values in units of
mag~day$^{-1}$.

\begin{figure}
\includegraphics[angle=0, width=8.5cm]{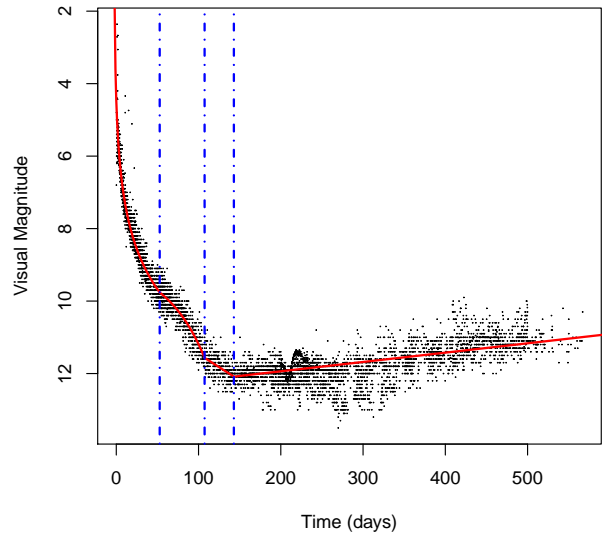}
\caption{Compiled declines from every outburst in Fig.~\ref{fig-outbursts},
  (first observed point to 500 days after peak) including the proposed
  1945 outburst, demonstrating the similarity of all outbursts. The
  red solid line indicates the best fit from the Bayesian analysis
  ($M_2$). Dot-dash blue lines indicate the break points, also from
  the Bayesian analysis.}
\label{fig-decline}
\end{figure}

The first observed data-point for the 1945 outburst was $\sim 68$~days
after the modelled beginning of the outburst. Given the fact that the
seasonal gap started $69.20$~days before the first observed point,
this means that there was an outburst only a few days after we entered
the seasonal gap. Especially for $\tau_2$, a $95\%$ credible interval
will give a value between $63.09$ and $69.09$ days. We were more lucky
with the 1985 outburst because before the first observed data-point
there was a seasonal gap of $65.40$ days. Since the modal value for
the start of the 1985 outburst $\widehat{\tau}_5 = 0.01$ days, this
means that this outburst started just after the end of the seasonal
gap. All the other outbursts did not seem to have any observational
gaps before the first observed data-points, giving us the chance to
observe them almost at the beginning. Fig.~\ref{fig-decline} shows all
the outburst data aligned by start date. The red solid line depicts
the chosen model ($M_2$) , whereas the blue dashed lines show the
breakpoints as the modal values $\widehat{t}_1$, $\widehat{t}_2$, and
$\widehat{t}_3$.

\begin{table*}
\caption{Summary of the posterior inference for the $M_2$
  model. Parameter $\tau_2$ is the start point of the 1945
  outburst.All times are in days since the modelled start points.}
\label{M_1:estimation}
\begin{center}
\begin{tabular}{ccccccc}     
  \hline                   
  & mean & mode & s.d. & $2.5\%$ & $50\%$ & $97.5\%$ \\
  \hline
$\gamma_1$ & 2.89 & 2.87 & 0.09 & 2.70 & 2.90 & 3.07 \\
$\alpha_1$ & 1.71 & 1.72 & 0.03 & 1.66 & 1.71 & 1.76 \\
$\beta_1$ & 2.53 & 2.53 & 0.18 & 2.19 & 2.52 & 2.91 \\
$t_1$ & 53.07 & 52.79 & 2.61 & 48.77 & 52.75 & 59.13 \\
$\alpha_2$ & 0.25 & 0.29 & 0.09 & 0.10 & 0.26 & 0.42 \\
$\beta_2$ & 0.02 & 0.02 & $3.18 \times 10^{-3}$ & $1.80 \times 10^{-2}$ & $2.18 \times 10^{-2}$ & $3.01 \times 10^{-2}$ \\
$t_2$ & 106.00 & 107.36 & 1.77 & 102.01 & 106.10 & 109.12 \\
$\alpha_3$ & 0.02 & 0.01 & $1.65 \times 10^{-3}$ & $1.22 \times 10^{-2}$ & $1.51 \times 10^{-2}$ & $1.87 \times 10^{-2}$ \\
$t_3$ & 142.89 & 142.95 & 1.79 & 139.40 & 142.91 & 146.41 \\
$\alpha_4$ & -2.55$\times 10^{-3}$ & -2.57$\times 10^{-3}$ & $5.06 \times 10^{-5}$ & $-2.65 \times 10^{-3}$ & $-2.55 \times 10^{-3}$ & $-2.45 \times 10^{-3}$ \\
$\tau_1$ & 3.45 & 3.47 & 0.24 & 2.98 & 3.45 & 3.93 \\
$\tau_2$ & 67.01 & 67.97 & 1.63 & 63.09 & 67.32 & 69.09 \\
$\tau_3$ & 0.62 & 0.66 & 0.07 & 0.42 & 0.64 & 0.70 \\
$\tau_4$ & 0.38 & 0.41 & 0.15 & 0.08 & 0.38 & 0.67 \\
$\tau_5$ & 0.09 & 0.01 & 0.08 & $2.45 \times 10^{-3}$ & 0.06 & 0.29 \\
$\tau_6$ & 0.45 & 0.49 & 0.10 & 0.24 & 0.46 & 0.65 \\
  \hline
\end{tabular}
\end{center}
\end{table*}

From this we have demonstrated that the 1945 data are consistent with
an outburst beginning and completing most of its decline during the
seasonal gap, but with the final decline and recovery visible. It does
not prove that this is an outburst, as we included it in our
analysis. But it is the simplest explanation of the data. As noted by
\citet{Oppenheimer1993} the data that are available show a form that
is only seen after outbursts. If we accept that there was an outburst
at this time, the mean recurrence interval goes down. Between 1933 and
1967 the mean interval was just over 10~years, compared with a mean
over all seven outbursts of approximately 15~years.


\section{Wavelet analysis}
\label{sec-wavelet}

We have subjected the lightcurve in Fig.~\ref{fig-lc} to a wavelet
analysis using the techniques of \citet{Torrence1998}. We applied the
implementation of these techniques in the Fortran~77 software
available from those authors available via
atoc.colorado.edu/research/wavelets. In order to apply a wavelet
analysis the data must be evenly sampled. Thus we have re-sampled to a
one day interval. Where a day had more than one observation, we have
taken an average, while we take a linear interpolation from the
adjacent dates over days with no data. Seasonal gaps are replaced with
the mean value for the entire lightcurve. As 90\% of the data are from
the quiescent periods the effect of outbursts on the mean is small. As
the analysis relies on a range of input data to generate each point in
the output data, we pad beyond the start and end of the data with the
mean value, until the total number of data points reaches the next
power of two. While we used a Morlet mother function
\citep{Torrence1998} that is designed to identify periodic signals in
the data, power appears in the resultant power plot due to
discontinuities in the data such as those seen at outburst.

While a complex wavelet function, such as used here, can allow the
separation of both phase and amplitude of the data and is useful for
identifying oscillations, a real wavelet function merges these two in
one component and can be helpful to isolate peaks or discontinuities
\citep{Meyers1993}. In this paper the wavelet power spectrum is
defined as the sum of the squares of the real and imaginary parts of
the wavelet transform \citep{Torrence1998}. Formally there is a cone
of influence (COI) outside which there is not enough information to
determine if a signal is real. Primarily this excludes periods longer
than the time--span of the data, but early and late in the time series
it is also affected by the padding, hence the cone--like shape.More
specifically, the wavelet power is expected to drop as we approach the
edge of the data, due to the infinite support of the Morlet wavelet
function.


Fig.~\ref{fig-power}a depicts the wavelet power plot of the optical
light curve in Fig.~\ref{fig-lc} following re-sampling to a one day
interval and interpolation of missing data as described above. The
vertical dotted lines represent the beginning of an outburst as
determined from our parametrisation approach. All the peaks in the
wavelet power spectrum are associated with an outburst. A similar
feature can be located for the year 1945, where
\citet{Oppenheimer1993} suggested an additional outburst. However, the
wavelet power is not as strong as all the other outbursts due to the
lack of data during the brightest part of the outburst (as we found in
Section~\ref{sec-decline} if an outburst had indeed occurred, it
started during the seasonal gap).

The most prominent peak in the wavelet power spectrum of
Fig.~\ref{fig-power}a is linked with the 1967 outburst. This has to
do with the fact that during the last stage of the magnitude decline
of the outburst the magnitude becomes fainter than 12, which might be
an indication that the 1967 outburst was more powerful than the others
observed. Note that the features we observe are not symmetric about
the beginning of each outburst. This asymmetry is because the wavelet
power is influenced by the data before and after the outbursts. The
signal for the 2006 outburst seems to be weaker than the rest of the
outbursts because part of it is outside of the COI. This means
that the wavelet power at larger scales will decrease as we approach
the edges because we pad the end of the time series with the mean
value.

\subsection{Pre--outburst Signal}
\label{ssec-preoutburst}

\begin{figure*}
\includegraphics[angle=0, width=18.5cm]{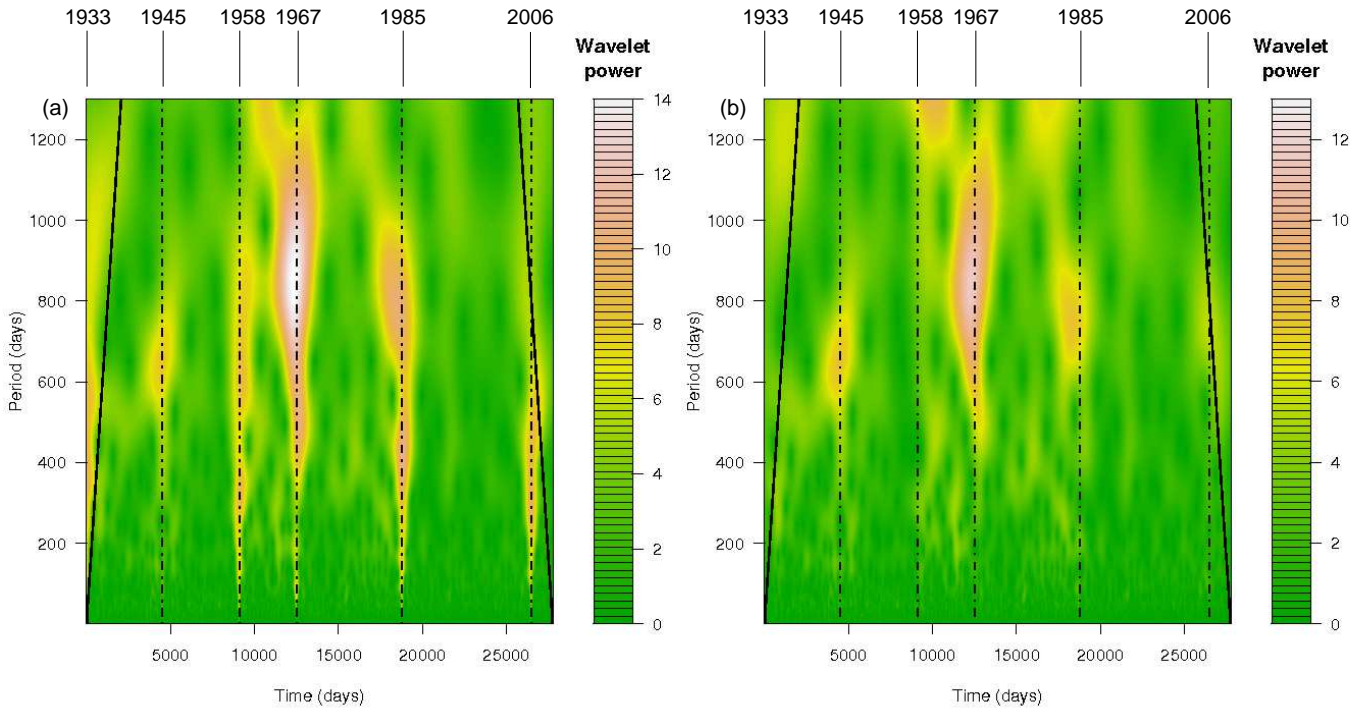}
\caption{Wavelet power plot of the lightcurve in Fig.~\ref{fig-lc}, (a)
  with the outbursts retained and (b) after replacing the outburst data
  with the mean of the data set.  Outburst dates are marked. Note that
  the pre-outburst signal remains even in the absence of the outburst
  discontinuity, and is visible for the 1945 obscured outburst. The
  scale of the power is different in (a) and (b), as indicated by the
  scale bars in each case.}
\label{fig-power2}
\label{fig-power}
\end{figure*}

A natural response of the wavelet power to discontinuities in the data
(such as those seen at outburst in Fig.~\ref{fig-power}a) is a peak in
the power at the time of the discontinuity.  We decided to remove the
first 143~days of each outburst and undertake the same analysis. The
removed data have been replaced with the mean of the time series, as
presented in Fig.~\ref{fig-power2}b. We tested the robustness of the
results for different choices for replacement data, including a
sinusoid of 50~day period and the residuals of subtracting the model
determined in Section~\ref{sec-decline}. Introducing different
artificial values for the removed data did not affect the outcome of
this analysis other than to insignificantly change the detailed
distribution of power in the wavelet power spectrum.

Fig.~\ref{fig-power2}b shows the resultant wavelet power
spectrum. There are peaks in the wavelet power spectrum at a period
range between 600 and 800~days that are associated with each of the
1945, 1985 and 2006 pre-outburst phases. This implies that these three
outbursts are most similar regarding the light curve. The fact that
part of the 2006 peak of the wavelet power is outside of the COI does
not suggest that the wavelet power is completely wrong. Since we pad
with the mean value of the time series at the edges, we expect the
wavelet power near the edges at large scales to be less than it
otherwise would be. This means that if we had more years of data, then
the 2006 wavelet power peak would be entirely inside the COI and
appear with enhanced wavelet power. The 1967 outburst is associated
with a feature in the period range 500 and 1100~days, whereas the 1958
outburst does not give any significant peak associated with the
pre--outburst. Thus we believe that the peak for 1958 in the wavelet
power spectrum in Fig.~\ref{fig-lc}b, when the outburst data is
retained,is due only to the data discontinuity. It is clear that all
of the other outbursts (apart from the 1958) are linked with a
specific form of wavelet power variation roughly around a period of
700--800~days which appears before the outburst begins, and so cannot
be due to the discontinuity.  Last but not least, the pre-outburst
phase and the post-outburst phase (i.e. after 143~days from the
outburst) of the light curve seem to follow the same pattern in all
cases.

This is another indication why we should remove the outbursts from the
analysis, as the peak for the wavelet power for the 1958 outburst in
Fig.\ref{fig-power}a is created only by the data
discontinuity. Therefore, the 1958 and 1967 outbursts might have
progressed differently and may provide a key to the outburst
mechanisms. In addition the 1958 outburst was only 9 years before the
1967 outburst, and also the 1958 outburst was not associated with any
fluctuation around that range.

To conclude, each outburst is associated with a peak in the wavelet
power spectrum around period 500 to 1000~days (apart from the 1958
outburst), which never appears outside of an outburst. A detailed
analysis shows that these peaks are part of the true signal and not
due to the observational uncertainty.

\section{Discussion}
\label{sec-discussion}

From our Bayesian analysis, using a piecewise model curve fitting for
the declines from outburst, we find breakpoints in the decline at
$\sim$53, $\sim$107 and $\sim$143~days after the outburst. It is not
clear that these are associated with any other changes in the system's
behaviour. The X-ray flux shows a change in the power-law of the
decline at around day~70 \citep{Bode2008,Ness2008} but there is no
obvious physical reason to relate this to changes in the optical light
curve. The super-soft phase also settled into a reasonably stable
state around day~50 \citep{Hachisu2007} but again this cannot be
clearly associated with the optical break by any physical mechanism.

It was found that the first observed point of the 1945 outburst was
$\sim$68~days after the actual beginning of the outburst, which means
that there was probably an outburst beginning at most a few days after
the seasonal gap began.  Using wavelet analysis the pre- and
post-outburst features in the light curve at this time can be
distinguished, and are similar to all the other outbursts (apart from
the 1958 one). Identifying an outburst in 1945 reduces the mean
interval outburst recurrence rate from $\sim$20 years
\citep{Starrfield2006} to $\sim$12.5 years. This also makes the nine
year interval between the 1958 and 1967 outbursts less anomalous -
moving it much closer to the median interval than when the 1945
outburst is excluded (from 1.25 standard deviations to 0.7 standard
deviations).

A more detailed look at the wavelet power for each outburst suggests
that the 1945, 1985 and 2006 outbursts had similar magnitude wavelet
peaks. The 1967 outburst has an unusually strong wavelet peak while in
1958 there is no significant peak. While we can say nothing about the
outburst mechanisms from this, it does suggest some observable
pre--outburst activity in most cases -- with the only analysed
outburst without such a peak followed by a second outburst within
9~years, still the shortest observed interval. The only major
parameter in outburst models \citep[e.g.][]{Yaron2005} that can be
varied on such short timescales is the mass accretion rate onto the
WD. Variable mass transfer also seems to be the only plausible process
that could generate the pre--outburst activity indicated by our
wavelet analysis. We note that the pre--outburst signal in the power
plots are similar regardless if we include or exclude the outbursts
themselves in the light curve (compare Figs.~\ref{fig-power}a and
\ref{fig-power2}b).


Since we know what a typical characteristic of each outburst is, then
we can use it for outburst prediction. Even 200~days before the
outburst the wavelet analysis shows this pre-outburst trend of the
star. For 455~days (an orbital period) this feature is less
significant but in some cases (1967 and 1985 outbursts) is still
recoverable. Hence, we have a means to predict outbursts in this
object. Further work will extend this to similar systems. If it turns
out to be robust then it may also have implications for the outburst
mechanism as a TNR should have no pre-outburst signal.

\subsection{Outbursts prediction}
\label{ssec-predict}

Since (almost) every analysed outburst is linked with a pre-outburst
feature in the wavelet power spectrum, this can be a useful tool for
outburst prediction. To test this, wavelet analysis for the time
interval between two outbursts has been applied. The starting point of
each time series was 500~days after the beginning of the outburst. The
end point of each time series was 455~days before the beginning of the
next outburst, which coincides with one orbital period. The aim here
was to investigate the appearance of the pre--outburst signal as we
approach outburst.

\begin{figure}
\includegraphics[angle=0, width=9.5cm]{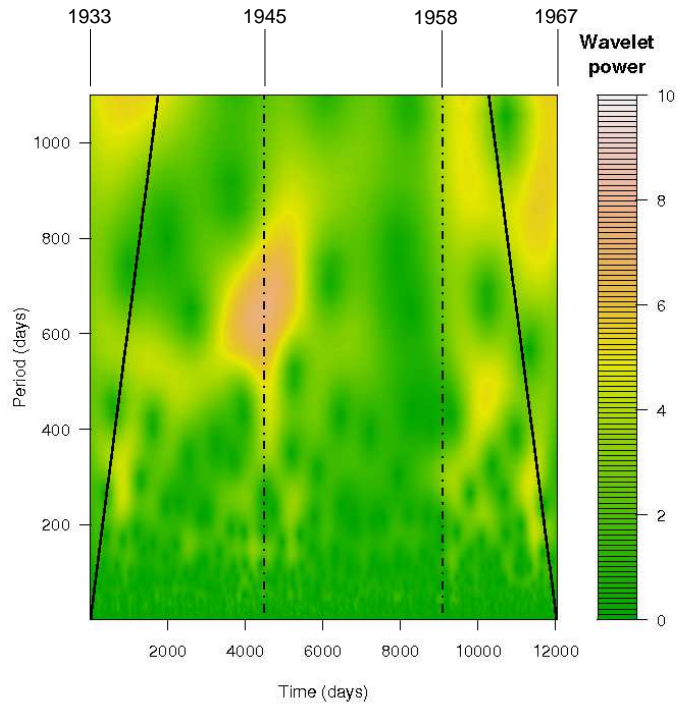}
\caption{Wavelet power plot restricted to the time period between 1933
and 1967, showing the outburst feature for the 1945 outburst which was
largely obscured by the seasonal gap.}
\label{fig-power4}
\end{figure}

In Fig.~\ref{fig-power4}, data values between the 1933 and the 1967
outbursts were analyzed. We can clearly see the peak in the wavelet
power spectrum at a period around 600--800~days associated with the
start of the 1945 outburst determined in Section~\ref{sec-decline}. No
such signal is apparent for 1958, consistent with the result for the
entire data-set. The pre--outburst signal for the 1967 outburst is
growing, albeit outside the COI. A similar analysis of the period from
1958 to 2006 showed similar signals for each outburst, with the 1967
one being by far the strongest.

The wavelet power of this peak drops as we approach the end of the
time series because more wavelet function is convolved with the
artificial data we pad the end of the time series with. The
interaction of the true data with the artificial data will reduce the
value of the wavelet power, the structure of which can only be
influenced by changes in the real data. Although this feature is
initially outside of the COI, it can still be recognized. As no other
feature like this at that period range can be observed in the rest of
the time series, this is consistent with the appearance of such a
feature at the most recent end of the wavelet power plot indicating
an approaching outburst. When we assumed we did not know that at the
end of these data there was an outburst, and did the same analysis up
to: i) 1 day before the outburst, ii) 100~days before the outburst,
and iii) 200~days before the outburst, we still see this feature for
every outburst where one was seen when the data from either side of
the outburst were included (1945, 1967, 1985 and 2006). The closer we
get to the outburst the better defined this feature is. Similar
analysis has been applied to all the time intervals between each pair
of outbursts. The outcome was that a peak in the wavelet power
spectrum like the one in Fig.~\ref{fig-power4} was observed before
every outburst, apart from the 1958 outburst. Given that this also had
the shortest interval to the next outburst in 1967 (which was then the
strongest outburst seen so far) it may be that something peculiar
happened at that date compared with the other outbursts. We do not
have data before 1933 at sufficient density to subject to this
analysis, so we cannot say anything about that outburst or the one in
1898.

\section{Conclusion}
\label{sec-conclusion}

We have analysed 76~years of optical photometry of the recurrent nova
RS~Ophiuchi, allowing analysis of both the quiescent and outburst
phases over that time. Bayesian analysis of the decline from maximum
of the combined data from the five accepted outbursts in the data and the
proposed 1945 outburst show that there is a preferred set of break
points when the form of the decline changes, at $\sim$53~days,
$\sim$107~days and $\sim$143~days after the outburst peak, after which
the optical light recovers towards the quiescent level over another
$\sim$65~days. With this empirical model of the decline, we are able
to place the start point of the 1945 outburst proposed by
\citet{Oppenheimer1993} as only about a day after the start of the
preceding seasonal gap. The 1945 data contribute so little to the
Bayesian analysis we can be confident that the decline seen here, but
no where else in the lightcurve outside of a confirmed outburst, is
due to a 7th outburst as proposed by \citet{Oppenheimer1993}.

Turning to the overall lightcurve, including the quiescent phases, a
wavelet analysis shows that there is a signal in wavelet power due to
the outbursts. While some of this signal is contributed by the
discontinuity due to the outburst, replacing the outburst data with
continuous data essentially indistinguishable from quiescent data only
reduces the strength of the signal prior to the known start dates for
the outbursts. Thus it appears that we have identified a pre--outburst
signature up to 450~days before outburst in the wavelet analysis,
which was apparent in all but the 1958 events. This holds out the
possibility of having an early warning of outbursts in the future. It
also suggests some implication of the accretion process in the onset
of outburst, as a TNR would not provide such a pre--outburst
signal. We are working to extend this work to similar objects.

\section*{Acknowledgments}

\noindent We thank Dr Mike Marsh for creating an independent plot
equivalent to Fig.~\ref{fig-power} as part of checking the outcomes
for this paper. The wavelet analysis was applied using the wavelet
software of Torrence and Compo (available at
atoc.colorado.edu/research/wavelets). We acknowledge with thanks the
variable star observations from the AAVSO International Database
contributed by observers worldwide and used in this research. SA was
supported by a PhD studentship from the University of Central
Lancashire and the Science \& Technology Facilities Council.

\bsp

\label{lastpage}
\end{document}